\documentclass[twocolumn]{revtex4-1}
\usepackage{graphicx}
\usepackage{amsmath}
\usepackage{MnSymbol}
\usepackage{mathrsfs}
\begin{document}
\title{Quantum Optical State Comparison Amplifier}
\author{Electra Eleftheriadou$^1$, Stephen M. Barnett$^2$ and John Jeffers$^{1*}$}

\affiliation{$^1$ Department of Physics, University of Strathclyde, John Anderson Building, 107 Rottenrow, Glasgow G4 0NG, U.K.\\
$^2$ School of Physics and Astronomy, University of Glasgow, Kelvin Building, University Avenue, Glasgow, G12 8QQ, U.K.\\
$^*$john.jeffers@strath.ac.uk}

\date{\today}
\begin{abstract}
It is a fundamental principle of quantum theory that an unknown state cannot be copied or, as a consequence, an unknown optical signal cannot be amplified deterministically and perfectly. Here we describe a protocol that provides nondeterministic quantum optical amplification in the coherent state basis with high gain, high fidelity and which does not use quantum resources. The scheme is based on two mature quantum optical technologies, coherent state comparison and photon subtraction. The method compares favourably with all previous nondeterministic amplifiers in terms of fidelity and success probability.
\end{abstract}
\maketitle

Signal amplification is a simple concept in classical physics. In electromagnetism there is no theoretical impediment to amplifying a time-varying electric field $E$ to form a perfectly-copied larger signal $gE$, where $g (>1)$ is the gain factor. There are, of course, practical limitations: the gain is typically saturated because only a finite amount of energy is available for the amplifier. Furthermore the utility of the device is limited in practice by the fact that the amplifier normally adds noise to the signal. 

A perfect quantum optical amplifier would increase the coherent amplitude of a state multiplicatively. For example, it would transform the coherent state $|\alpha\rangle$, the nearest quantum equivalent to a classical stable wave, as follows
\begin{eqnarray}
|\alpha \rangle \rightarrow |g\alpha \rangle.
\end{eqnarray}
Quantum-level linear optical amplifiers have stringent limitations on their operation. It is impossible to amplify an unknown quantum optical signal without adding noise \cite{haus}, the minimum value of which is a consequence of the uncertainty principle \cite{caves}. This extra required added noise swamps the quantum properties of a signal. Were it otherwise it would be possible to violate the no-cloning theorem \cite{wootters} and achieve superluminal communication \cite{herbert}.

Ralph and Lund \cite{ralph1} suggested that this noise limit could be beaten by nondeterministic amplifiers: ones that work only in postselection. Such amplifiers transform the coherent state $|\alpha \rangle \rightarrow  c |g\alpha \rangle$, 
where $c$ satisfies $|c| \leq \frac{1}{g}$.
They proposed an amplifier based on the quantum scissors device \cite{scissors1,scissors2} and this was later realised experimentally \cite{ralph2,grangier}. The scheme has been extended to amplification of photonic polarization qubits using two such amplifiers \cite{swiss,kocsis}. 

Unfortunately the amplifier uses single photons as a resource, its success probability is only a few percent and it works only in the $|0\rangle$ and $|1\rangle$ basis, so the condition on the output is $|g \alpha| \ll 1$. This restriction could be circumvented by operating several amplifiers in parallel \cite{ralph1}, or using a two-photon version of the quantum scissors device \cite{jeffers}, but the requirement of multiple coincident separately-heralded photons renders the effective probability of amplification tiny. Photon addition \cite{walker} and subtraction \cite{wenger} can also be used to form a nondeterministic amplifier with $g=\sqrt{2}$ \cite{addsubtract1,addsubtract2}, but with the same type of limitations as the scissors-based device. Another scheme relies on weak measurements and the cross-Kerr effect and so also has low success probability \cite{croke}. 

A protocol for state amplification without using quantum resources was suggested by Marek and Filip \cite{marekfilip}. Surprisingly, thermal noise is added to the signal. The state is then Bose-conditioned by photon subtraction, which performs the effective amplification. The amplification is not perfect, but it produces larger amplitude output states with phase variances smaller than those of the input \cite{usuga}. The scheme can be improved slightly if a standard optical amplifier is used to add the initial noise \cite{jj2,laterguy}. The success probability in all cases is limited by the photon subtraction probability, of the order of a few percent.

Here we describe a remarkably simple method for amplifying coherent states based on comparing the input state with a known coherent state \cite{andersson}. This type of comparison has already been used in an experimental realisation of a quantum digital signature scheme \cite{qds} and in various binary quantum receivers \cite{kennedy, dolinar, bondurant}. As an amplifier it has several advantages over earlier methods. 

We assume that Alice sends coherent states selected at random from a known set to Bob. Bob's task is to amplify them, e.g. for later splitting and distribution of identical copies, or to determine the phase of the coherent state accurately. Higher amplitude coherent states have a smaller phase variance, so this amounts to the sharing of a reference frame - an important task in quantum communication \cite{reframe}. There are many other possibilities.

\begin{figure}
\includegraphics[width=0.9\columnwidth]{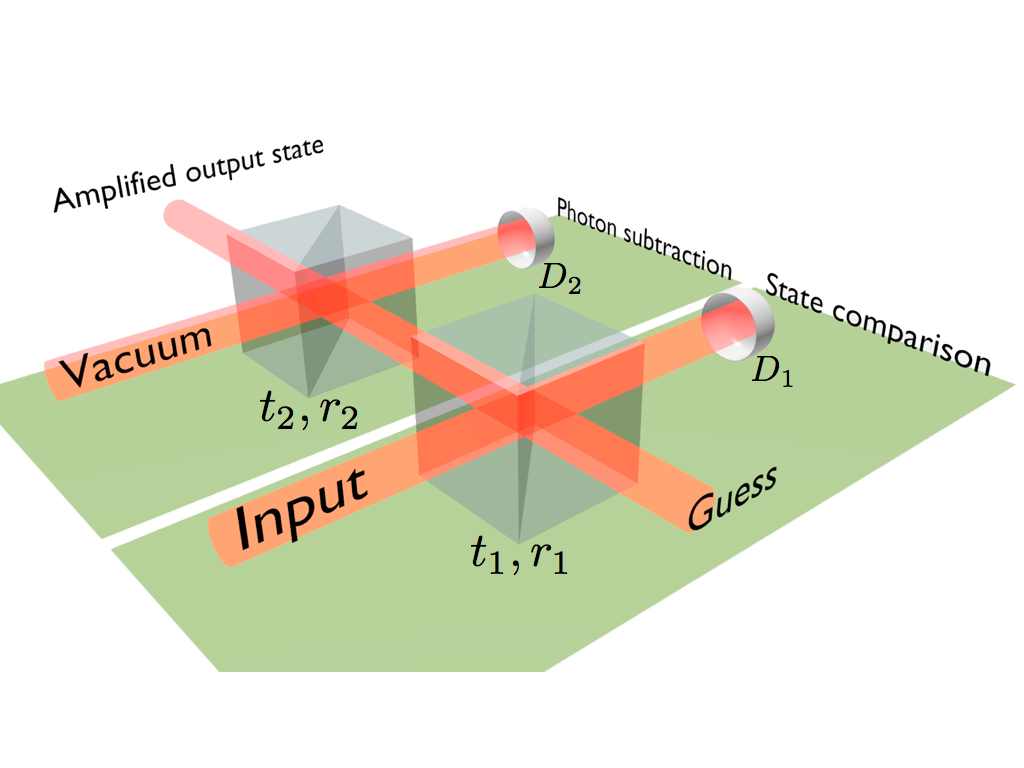}
\caption{(Colour online) The state comparison amplifier. Bob attempts to null Alice's input with his guess state at the first beam splitter. The second beam splitter and detector are used for photon subtraction. The output state is accepted if the first detector does not fire and the second one does.}
\label{fig1}
\end{figure}

Bob performs the amplification using the device shown in Fig. \ref{fig1}. He mixes the unknown input state with another coherent state (the guess state) at a beam splitter and compares them. The beam splitter has transmission and reflection coefficients $t_1$ and $r_1$ which we take to be real and there is a phase change of $\pi$ on reflection from the lower arm. One output arm of the beam splitter falls upon a photodetector and the other output is postselected based on no photocounts being recorded. This system performs a state comparison between the reflected part of the guess state and the transmitted part of the input state. 
If Alice's input state is $|\alpha \rangle$ and Bob chooses his guess state to be $|\beta \rangle=|t_1 \alpha/r_1 \rangle$ (correctly) the transmitted input state interferes destructively with the reflected guess state and the detector cannot fire.  The output in the upper arm is then a coherent state of amplitude $\alpha/r_1$, which is larger than $\alpha$. This is the gain mechanism for the device. If Bob chooses his guess state incorrectly, coherent light leaks into the detector arm. This can also sometimes cause no counts at the detector. Thus the output state, \textit{conditioned} on the detector not firing, is generally a mixed state, weighted by the probability that no counts are recorded. The probability is maximised when destructive interference occurs in the detector arm, for which the output coherent amplitude also reaches its maximum. 

Bob can improve the quality of his output at small cost to the success probability if he performs a photon subtraction on the output. Coherent states are eigenstates of the annihilation operator and so subtraction has no effect on them \cite{zavatta}, but for a mixture of coherent states with different mean photon numbers the probabilities in the mixture are adjusted. When the detector fires, a subtraction occurs and it is more likely to have been due to a high-amplitude coherent state rather than a low amplitude one. Thus a subtraction is more likely when Bob has chosen his guess state well. If the subtraction is performed with a beam splitter of transmission coefficient $t_2$, then $g=t_2/r_1$ is the nominal gain of the composite system.

For input and guess states $|\alpha \rangle$ and $| \beta \rangle$ the coherent amplitude in the nominal vacuum output is $t_1 \alpha - r_1 \beta$ and the other beam splitter output passes to the subtraction stage. The amplitude in the subtraction arm is therefore 
$-r_2(t_1 \beta+r_1 \alpha)$ and the output amplitude is $t_2(t_1 \beta+r_1 \alpha)$. We assume that the input and guess states are chosen from probability distributions over the coherent states
\begin{eqnarray}
\label{inputstates}
\nonumber
\hat{\rho}_{\mbox{in}} &=& \int d^2 \bar{\alpha} P(\bar{\alpha})| \bar{\alpha} \rangle \langle \bar{\alpha} |, \\
\hat{\rho}_{\mbox{g}} &=& \int d^2 \bar{\beta} Q(\bar{\beta})| \bar{\beta} \rangle \langle \bar{\beta} |,
\end{eqnarray}
and calculate the output state and the fidelity based on these and the properties of the device. The fidelity is 
\begin {eqnarray}
F=\int d^2\alpha P(\alpha) \langle g \alpha |\hat{\rho}_{\mbox{out}} | g \alpha \rangle,
\end{eqnarray}
where $\hat{\rho}_{\mbox{out}}$ is the output state conditioned both on the input state distributions from eq.(\ref{inputstates}) and on the successful operation of the device. This is the probability that the output state passes a measurement test comparing it to the amplified version of the input state and can be written
\begin{eqnarray}
\label{fidform}
\nonumber F&=& P(T|S) = \frac{P(T,S)}{P(S)} \\
&=& \frac{\int d^2 \bar{\alpha} \int d^2 \bar{\beta} P(T|S,\bar{\alpha},  \bar{\beta}) P(S|\bar{\alpha},  \bar{\beta}) P(\bar{\alpha})Q(\bar{\beta})}{\int d^2 \bar{\alpha} \int d^2 \bar{\beta} P(S|\bar{\alpha},  \bar{\beta}) P(\bar{\alpha})Q(\bar{\beta})}
\end{eqnarray}
where $P(T|S)$ is the probability that the output state will pass the fidelity test given that the device operates successfully.  

\begin{figure}
\includegraphics[width=0.9\columnwidth]{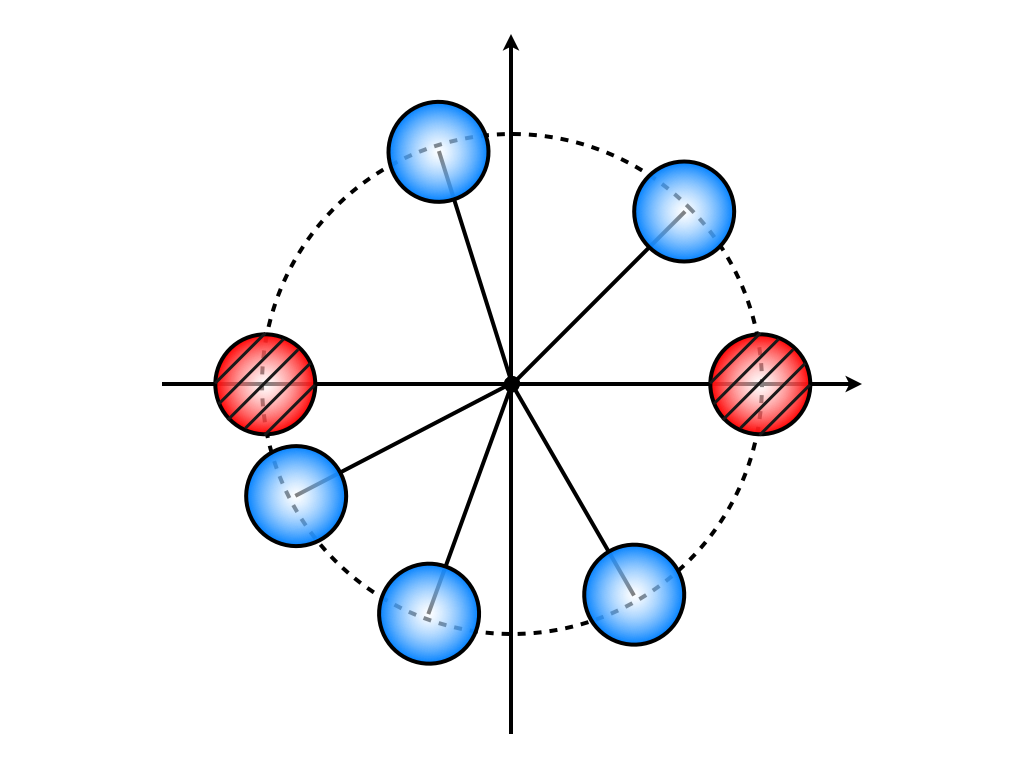}
\caption{(Colour online) The sets of input states considered, binary (red, hatched), phase-covariant (blue shaded).}
\label{fig2}
\end{figure}

So far we have made no assumptions about the forms of the input and guess probability distributions, but it is instructive to consider two cases in order to comply with the requirements of a realistic communication system. The first scenario is one in which the set of states to be amplified is restricted to the binary alphabet $\{ |\alpha \rangle, |-\alpha \rangle \}$ and in the second each state in the set has the same mean photon number, but with completely uncertain phase (Fig. \ref{fig2}). Here the mean photon number could either be agreed in advance, or determined by simply measuring the first few states. Without loss of generality we will assume from here on that $\alpha$ is real and positive.

Suppose that Alice chooses randomly from the binary set of states 
\begin{eqnarray}
P( \bar{\alpha}) = \frac{1}{2}\left[ \delta^2 (\bar{\alpha} - \alpha) + \delta^2 (\bar{\alpha} + \alpha) \right].
\end{eqnarray}
The best choice for Bob's guess state is to choose randomly from the set $\{ |\pm t_1 \alpha/r_1 \rangle\}$. Suppose, without loss of generality, that he chooses + so
$Q(\bar{\beta}) = \delta^2 (\bar{\beta} - t_1 \alpha / r_1)$. The probabilities which form the fidelity and success probability in eq. (\ref{fidform}) are straightforwardly calculated \cite{kelleykleiner} and we leave details to the supplemental material.
\begin{figure}
\includegraphics[width=0.9\columnwidth]{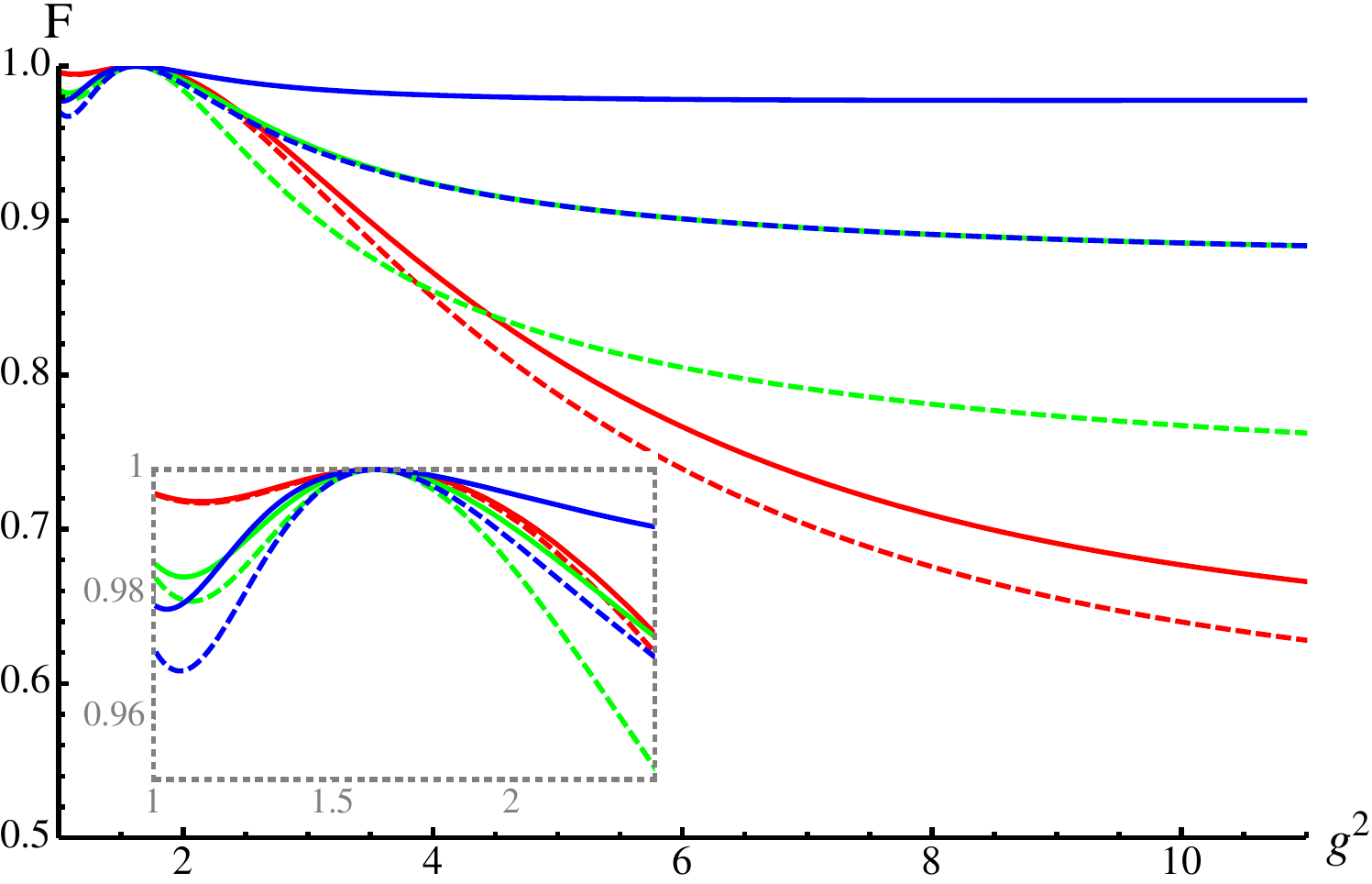}
\caption{(Colour online) Fidelity as a function of intensity gain for the binary system. The red (bottom two) , green and blue (top two) plots correspond to $\alpha^2= 0.1, 0.5, 1$. Full (dashed) curves are for detector quantum efficiencies of  $\eta =1 (0.5)$. The intensity reflection coefficient of the subtraction beam splitter is 0.1. The inset shows an expanded view of the low gain region. The fidelity for $\alpha^2 = 1, \eta = 0.5$ is very close to that for $\alpha^2 = 0.5, \eta = 1$ for gains greater than about 3.}
\label{fig3}
\end{figure}

Fig. \ref{fig3} shows the fidelity as a function of gain. The fidelity drops as the gain is increased initially, rises to unity at intermediate gain and then decays. 
The increase at intermediate gain is a combined effect of the photon subtraction and comparison. If the first beam splitter is 50/50, then its conditioned output that falls upon the subtraction beam splitter has a coherent amplitude of either $\sqrt{2} \alpha$ (if Bob has chosen his state correctly) or zero (incorrectly). If it is zero then no photon subtraction can take place. Therefore, if there is a photon subtraction Bob must have chosen correctly and a perfect amplified copy of the state is made. At high gain for perfect detector efficiency the fidelity decays to $1/(1+\exp{[-4\alpha^2]})$, which approaches unity for large $\alpha$. Fidelity is not the only possible measure of output quality and in the supplemental material we show that the amplifier has a noise figure which increases with gain.

\begin{figure}
\includegraphics[width=0.9\columnwidth]{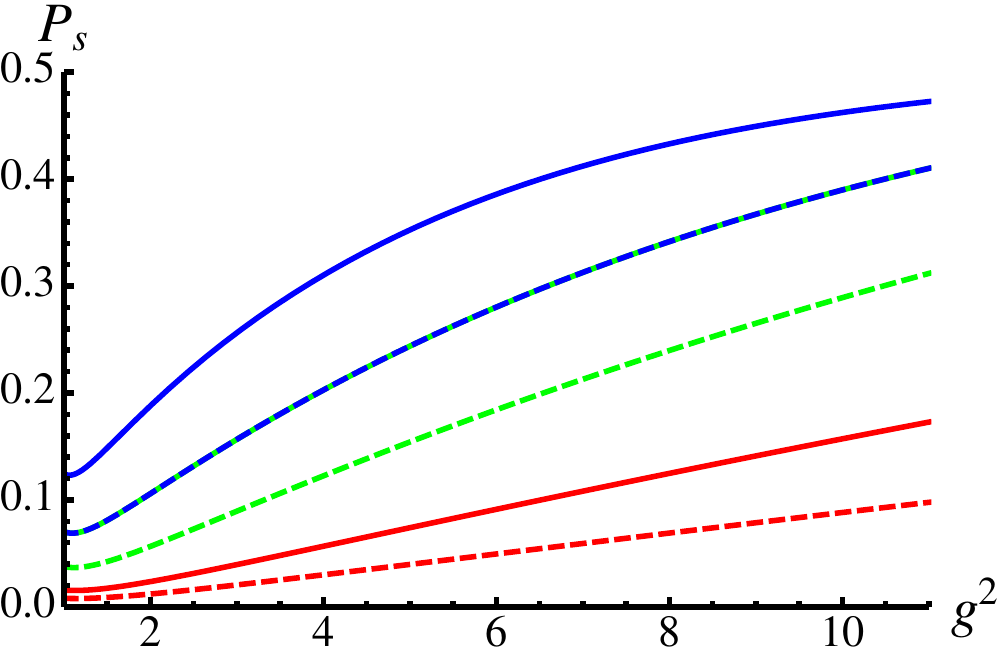}
\caption{(Colour online) Success probability as a function of intensity gain for the binary system. Parameters as for Fig. \ref{fig3}.}
\label{fig4}
\end{figure}
The success probability (Fig. \ref{fig4}) is dominated at low gain by the photon subtraction probability, but for higher gains the subtraction probability approaches one because the coherent amplitude is large.

It is important to note that the performance of our device is relatively insensitive to experimental detector imperfections. The fidelity is reasonably robust to nonunit quantum efficiency, as shown in Fig. \ref{fig3}. At high gain a nonunit quantum efficiency reduces the fidelity to that which would be obtained by operating the device with reduced coherent amplitude input $\alpha \sqrt{\eta}$. The effects on success probability shown in Fig. \ref{fig4} depend on two competing factors. There is an increased probability of obtaining no counts at the comparison detector, but a decreased probability of photon subtraction. In a realistic experimental scenario the ideal is to keep dark count rates low enough that they do not affect the results. For example, in the state comparison experiment of reference \cite{qds} the photon flux at the detectors is of the order of $10^7s^{-1}$ and the dark count rates for the SPAD detectors are 320$s^{-1}$, rendering the effects of dark counts insignificant.

If Bob and Alice do not initially share a phase reference and all that Bob knows is the mean photon number of Alice's input then 
\begin{eqnarray}
P( \bar{\alpha}) = \frac{1}{2 \pi \alpha}\delta (|\bar{\alpha}| - \alpha),
\end{eqnarray}
with $Q(\bar{\beta})$ as before. The relevant probability calculations are again left for the supplemental material, together with plots of the fidelities and success probabilities. It is instructive to compare the fidelity with that obtained using other amplification methods. In Fig. \ref{fig5} we show the fidelity obtained using the state-comparison amplifier, using the quantum scissors \cite{ralph1,ralph2,grangier} and using the noise addition amplifier \cite{marekfilip,usuga} for $\alpha^2=0.5$.  The advantages of the state-comparison amplifier are obvious. We can see that for the binary alphabet the state comparison amplifier significantly outperforms the other systems. The effect of the photon subtraction ensures perfect amplification for twofold gain and no other amplifier can reach this. 

\begin{figure}
\includegraphics[width=0.9\columnwidth]{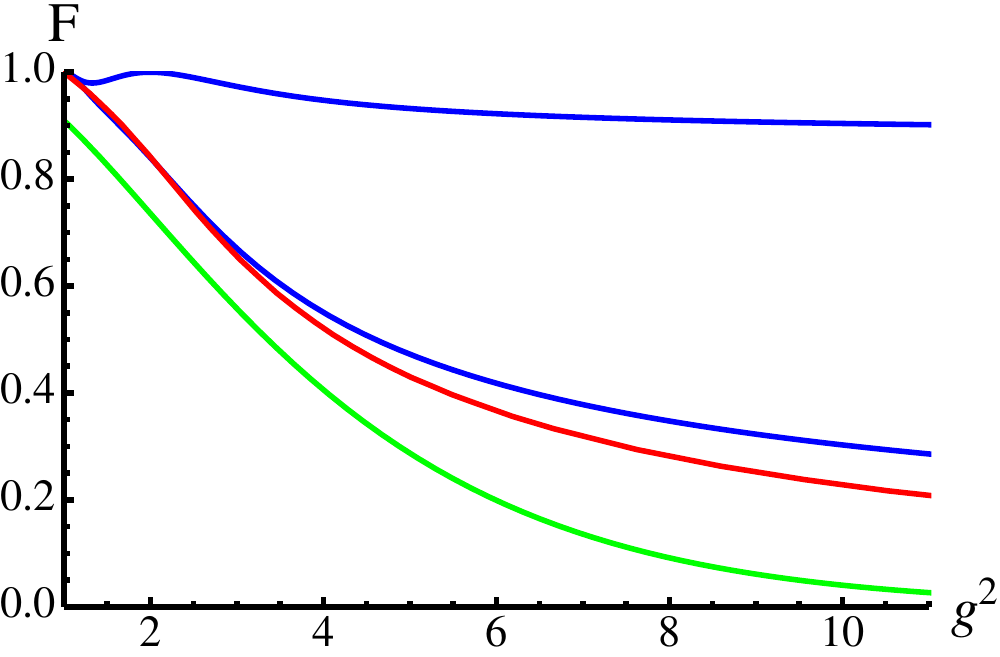}
\caption{(Colour online) Fidelity as a function of intensity gain for the state comparison amplifier. Curves are from top to bottom: state comparison binary system (blue upper), state comparison phase-covariant (blue lower), noise addition (red) and scissors-based (green). Detector efficiencies and photon subtractions are assumed perfect in all cases.}
\label{fig5}
\end{figure}

For the phase covariant state set, again the state comparison amplifier outperforms the other systems, although for low gain its edge over the noise addition amplifier is minimal in terms of fidelity. However, it does have another advantage over this system. When it works the state comparison amplifier provides knowledge of both the state to be amplified and the amplified state, knowledge which is not available in the noise addition amplifier. 

The lower fidelity associated with the scissors-based amplifier is largely due to the fact that it can only produce a superposition of zero and one photon, which is not useful for amplifying a state of mean photon number 0.5. For much lower mean input photon numbers the scissors-based amplifier has a higher fidelity than other methods.

It is clear that to be useful in future quantum communication systems postselecting amplifiers must approach the ultimate limits of performance \cite{pandey}. Here we have described a nondeterministic amplifier that outperforms other schemes over a wide range of input amplitudes and gains. It does not require quantum resources and operates with high fidelity and high success probability. It uses two already demonstrated experimental techniques and is relatively straightforward to implement. The gain can be chosen via the reflectivity of a beam splitter. 

The main reason why this amplifier works well is that it uses the available information about the input states effectively. Amplification is performed by dumping energy into the system - an optical mode. This works best if we do it in the appropriate basis, so if we want to amplify coherent states we should place the energy in the coherent state basis. 

The amplifier is robust to realistic values of detector imperfections. It still works well for nonunit detector efficiencies and dark count rates should be low enough to render them unimportant. Losses within the other components, such as at beam splitters or in connecting fibres, will be small and will reduce fidelity and/or gain by a commensurate amount.

The device works best in a limited state space, a feature common to all nondeterministic amplifiers, although the limits are different for each one. Both the scissors-based and photon addition/subtraction amplifiers have an output which cannot contain more than one photon and for the noise addition amplifier the gain is tailored to the input amplitude to maximize phase concentration. Our amplifier turns the state space limitation to an advantage, in that the amplifier can be tailored to work in the basis used in a particular communication system. However the state space that the device works on can always be widened at a cost to the success probability. Any cost to the fidelity depends on the gain chosen and the states added. We also remark that the amplifier described here provides gain that is dependent on the input state and this is the case for all postselecting amplifiers so far. It renders them effectively nonlinear, but it does not bring into question their status as amplifiers \cite{pandey}.

None of the earlier schemes can amplify superpositions of coherent states and the same is true of the state comparison amplifier as proposed here. In principle this device could amplify a limited set of superpositions of coherent states, but to do so it would require as inputs guess states that were themselves superpositions. However, superpositions are of limited use in communication systems, as propagation quickly destroys coherence. 

The most striking results of the state comparison amplifier are for an input chosen from a binary set of coherent states, where for a gain of just less than twofold perfect amplification can be achieved. This suggests an application of the device as an ideal quantum optical repeater, stationed every few km in a low-loss optical fibre communication system - the quantum equivalent of erbium doping. 

This work was supported the Royal Society, the Wolfson Foundation and the UK EPSRC.

\begin{widetext}
\section*{Supplemental Material}
\subsection*{Calculation of Fidelities and Success Probabilities for the Binary System}
The fidelities and success probability are given by Eq. (4) in the main paper and its denominator. They depend upon the probability that the device operates successfully given that pure coherent states $|\alpha \rangle$ and $| \beta \rangle$ are input by Alice and Bob. This is the probability that the first detector does not fire and the second one does, and is provided by the the Kelley-Kleiner formula \cite{kelleykleiner}
\begin{eqnarray}
P(S|\alpha, \beta) &=& \mbox{Tr} \left\{ \hat{\rho} : \exp{(-\eta_1\hat{a}_1^\dagger\hat{a}_1)}  \left[  1- \exp{(-\eta_2\hat{a}_2^\dagger\hat{a}_2)} \right] : \right\},
\end{eqnarray}
where $\hat{\rho}$ is the three mode output of the device, $\hat{a}_1$ is the annihilation operator for detector mode 1, and $\eta_1$ is the quantum efficiency of detector 1. As the input states are pure coherent states the outputs will also be pure coherent states, and
\begin{eqnarray}
\label{psab}
P(S|\alpha, \beta) &=&  \exp{(-\eta_1|t_1 \alpha - r_1 \beta |^2}) \left[  1- \exp{(-\eta_2 r_2^2 |t_1 \beta + r_1 \alpha|^2) } \right].
\end{eqnarray}
The output state given that the input amplitudes were $\alpha$ and $\beta$ is simply $|t_2(t_1\beta+r_1\alpha) \rangle$, and the overlap of this with the state $ |g \alpha \rangle = |t_2 \alpha/ r_1 \rangle$ is the probability that the output state passes the measurement test that it is an amplified version of the coherent input. 
Finally we can use these probability distributions in, together with the input probability distributions for the binary system to obtain the success probability and the fidelity
\begin{eqnarray}
\nonumber
P(S) &=& \frac{1}{2} 
\left( 1- \exp{ \left[ -\eta_2 g^2 \alpha^2 \left(1/t_2^2 -1  \right) \right] } + \exp{\left[-4 \eta_1 \alpha^2 \left( 1-t_2^2/g^2 \right)\right]} 
 \left\{ 1- \exp{\left[ -\eta_2 g^2 \alpha^2 \left(1/t_2^2 -1  \right) \left( 1- 2t_2^2/g^2 \right)^2\right]} \right\} \right) \\ & &\\
\nonumber
P(T,S)&=& \frac{1}{2} 
\left( \rule{0mm}{4mm} 1- \exp{ \left[ -\eta_2 g^2 \alpha^2 \left(1/t_2^2 -1  \right) \right] }  +  \exp{\left[-4 \eta_1 \alpha^2 \left( 1-t_2^2/g^2 \right)\right]} \right. \\  &\times& \left. \left\{ 1- \exp{\left[ -\eta_2 g^2 \alpha^2 \left(1/t_2^2 -1  \right) \left( 1- 2t_2^2/g^2 \right)^2\right]} \right\}  \exp{ \left[-4g^2 \alpha^2 \left( 1-t_2^2/g^2 \right)^2 \right] }\right).
\end{eqnarray}

\subsection{Noise Figure}

One of the more standard measures of amplifier quality is the signal to noise ratio (SNR) of the output, or the noise figure, which is the ratio of the output SNR to that of the input. These are quantities typically used in optical physics to quantify the quality of a device. We examine here a simple signal to noise ratio for the $x_1 = \left( \hat{a} + \hat{a}^\dagger \right)/\sqrt{2}$ quadrature,
\begin{eqnarray}
\mbox{SNR} =\frac { \langle x_1 \rangle} {\sqrt{(\Delta x_1)^2}}.
\end{eqnarray}
This quantity is appropriate for a nominally real coherent input $\alpha$, for which it takes the value $2|\alpha|$. We examine the binary alphabet case for which a relatively straightforward set of calculations find 
\begin{eqnarray}
\langle x_1 \rangle &=& \sqrt{2} g\alpha\left[ P(\alpha|S) +  \left( 1-2t_2^2/g^2 \right) P(-\alpha|S)\right],\\
\nonumber \langle x_1^2 \rangle &=& \frac{1}{2} \left\{\rule{0mm}{4mm} 1 +4g^2\alpha^2\left[ \rule{0mm}{4mm}P(\alpha|S) + \left( 1-2t_2^2/g^2 \right)^2 P(-\alpha|S)\right] \right\}, 
\end{eqnarray}
where the probabilities $P(\alpha|S)$ and $P(-\alpha|S)$ are the conditional probabilities that the input coherent amplitudes were $\alpha$ and $-\alpha$ respectively, given that the device operated successfully. 

We plot the noise figure based on these formulae in Fig. \ref{nf}. This clearly shows improvement for all gains larger than about 1.3. The noise figure is relatively insensitive to input coherent amplitude, and (not shown) to detector quantum efficiency. 
\begin{figure}
\includegraphics[width=0.65\columnwidth]{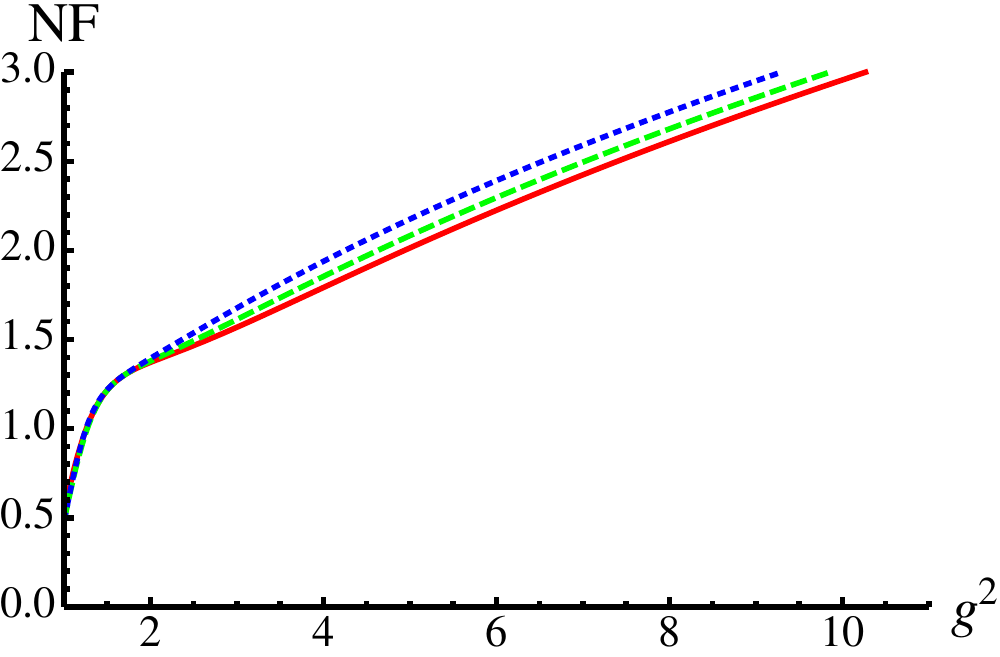}
\caption{(Colour online) Noise figure (NF) as a function of intensity gain for the binary system. Curves are from bottom to top (red, green dashed, blue dotted) $\alpha^2=0.1, 0.5, 1$. The intensity reflection coefficient of the subtraction beam splitter is 0.1.}
\label{nf}
\end{figure}

\subsection{Phase-covariant amplifier }
The  phase covariant amplifier calculations, for which Alice's input state probability distribution is \newline $P( \bar{\alpha}) = \frac{1}{2 \pi \alpha}\delta (|\bar{\alpha}| - \alpha)$, require a phase integral to be performed, 
\begin{eqnarray}
\nonumber P(S) &=& \int d^2\bar{\alpha}\int d^2\bar{\beta} P(\bar{\alpha}) Q(\bar{\beta})  P(S|\bar{\alpha}, \bar{\beta}) \\
\nonumber &=& \frac{1}{2 \pi} \int d\theta \exp{[-2 \eta_1 \alpha^2 t_1^2 (1-\cos \theta)]}  \left( 1-\exp{\{-\eta_2 \alpha^2 r_2^2 [1-2r_1^2(1-r_1^2)(1-\cos \theta )]/r_1^2\} } \right) \\
\nonumber &=& \exp{[-2 \eta_1 \alpha^2 (1-t_2^2/g^2)]} I_0[2 \eta_1 \alpha^2 (1-t_2^2/g^2)]\\
\nonumber &-& \exp{[-2 \eta_1 \alpha^2 (1-t_2^2/g^2) -\eta_2 \alpha^2 g^2 (1/t_2^2 -1) +2\eta_2 \alpha^2 (1-t_2^2) (1-t_2^2/g^2)]}  \\
&\times& I_0[2 \eta_1 \alpha^2 (1-t_2^2/g^2)  -2\eta_2 \alpha^2 (1-t_2^2) (1-t_2^2/g^2)], 
\end{eqnarray}
where $I_0$ is the modified Bessel Function of zero order. The numerator in the fidelity is the same function as above, but with the substitution
\begin{eqnarray}
\eta_1 \rightarrow \eta_1+ g^2-t_2^2.
\end{eqnarray}

We plot the fidelity and success probability in Figs. ({\ref{fig5}) and (\ref{fig6}). 
\begin{figure}
\includegraphics[width=0.65\columnwidth]{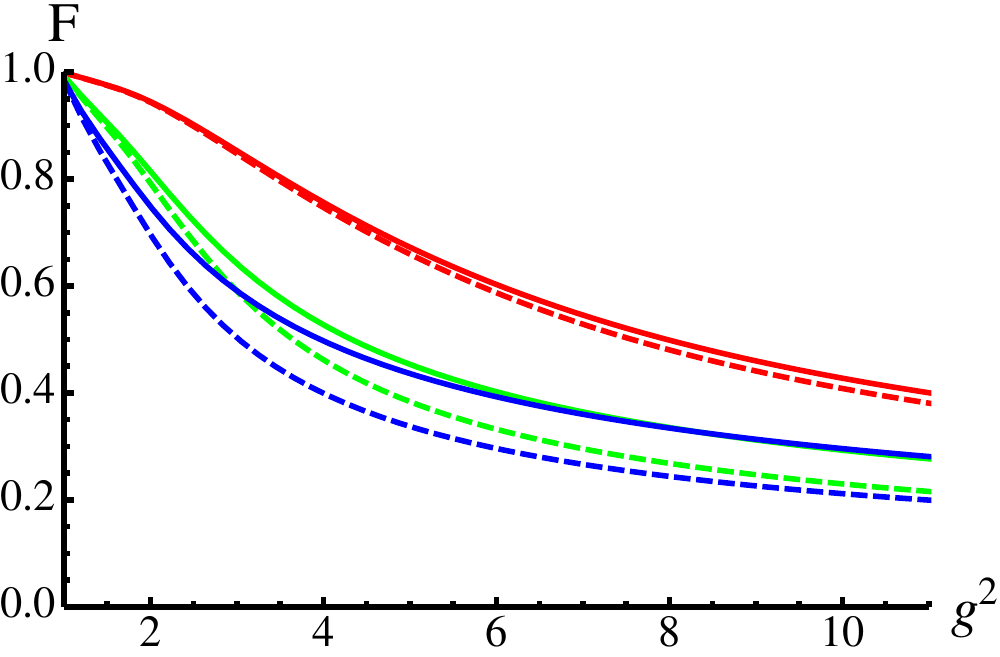}
\caption{(Colour online) Fidelity as a function of intensity gain for the phase-covariant system. Curves are from top to bottom (red, green, blue) $\alpha^2=0.1, 0.5, 1$. Full (dashed) curves are for $\eta=1 (0.5)$. The intensity reflection coefficient of the subtraction beam splitter is 0.05.}
\label{fig5}
\end{figure}
The fidelity decays with gain and the success probability increases with gain, as might be expected. The effect of photon subtraction is not as strong as in the binary case, but is still present. The other main difference is that the fidelity is highest for low mean photon number inputs, largely because even if the wrong state from the ring is amplified, it still has significant overlap with the nominal amplified state. This is not the case for higher amplitude input states, for which the amplified version of any state not sufficiently close to Bob's guess state is effectively orthogonal to the nominal amplified output. 

\begin{figure}
\includegraphics[width=0.65\columnwidth]{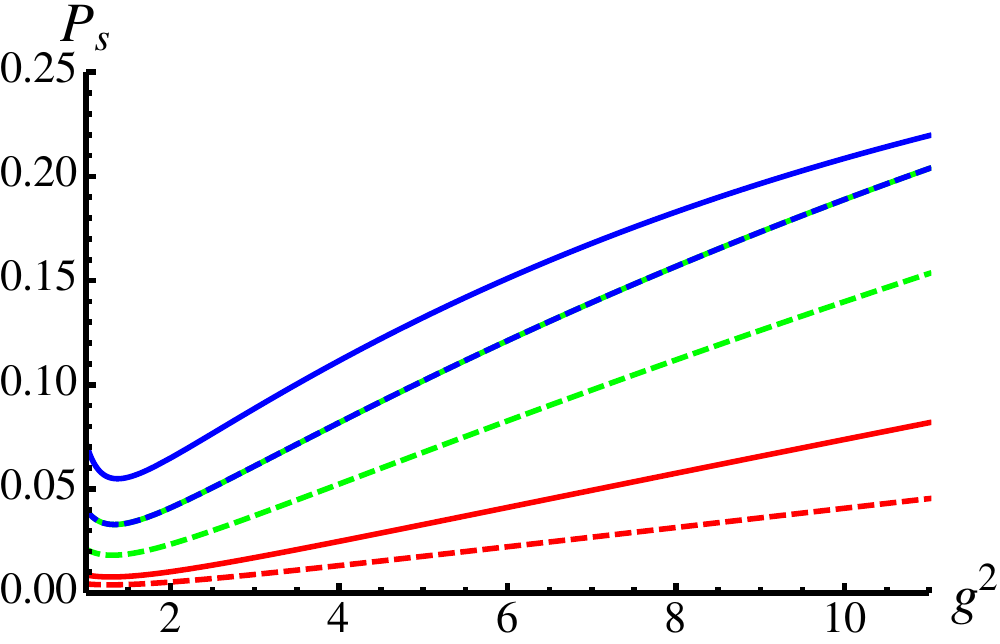}
\caption{(Colour online) Success probability as a function of intensity gain for the phase-covariant system system. Curves are from top to bottom (blue, green, red) $\alpha^2=1, 0.5, 0.1$. Full (dashed) curves are for $\eta=1 (0.5)$. The intensity reflection coefficient of the subtraction beam splitter is 0.05.}
\label{fig6}
\end{figure}

The success probability (Fig. \ref{fig6}) is of similar magnitude to the equivalent for the binary system.  It rises with gain, as increasing the gain amounts to adding more photons into the guess state beam splitter input port, and increasing the transmission of this beam splitter.
}
\end{widetext}
\end{document}